# Title

Sensitivity analysis for causality in observational studies for regulatory science

# Authors and Affiliations

Díaz, Iván[1]; Lee, Hana[2]; Kıcıman, Emre[3]; Schenck, Edward J;[4] Akacha, Mouna[5]; Follman, Dean[6]; Ghosh, Debashis[7]

# Corresponding author

Iván Díaz
Division of Biostatistics
Department of Population Health
New York University Grossman School of Medicine
180 Madison Ave, Office 2-52
New York, NY 10016
Ivan.diaz@nyu.edu

# Conflict of interest statement

# Disclaimer

The contents are those of the author(s) and do not necessarily represent the official views of, nor an endorsement by, FDA/HHS, or the U.S. Government.

# Abstract

Recognizing the importance of real-world data (RWD) for regulatory purposes, the United States (US) Congress passed the 21st Century Cures Act[1] mandating the development of Food and Drug Administration (FDA) guidance on regulatory use of real-world evidence. The Forum on the Integration of Observational and Randomized Data (FIORD) conducted a meeting bringing together various stakeholder groups to build consensus around best practices for the use of RWD to support regulatory science. Our companion paper[2] describes in detail the context and discussion carried out in the meeting, which includes a recommendation to use a *causal roadmap* for complete pre-specification of study designs using RWD. This article discusses one step of the roadmap: the specification of a procedure for sensitivity analysis, defined as a procedure for testing the robustness of substantive conclusions to violations of assumptions made in the causal roadmap. We include a worked-out example of a sensitivity analysis from a RWD study on the effectiveness of Nifurtimox in treating Chagas disease, as well as an overview of various methods available for sensitivity analysis in causal inference, emphasizing practical considerations on their use for regulatory purposes.

---

[1] Division of Biostatistics, Department of Population Health, New York University Grossman School of Medicine, New York, NY, USA.
[2] Office of Biostatistics, Office of Translational Sciences, Center for Drug Evaluation and Research, U.S. Food and Drug Administration
[3] Microsoft Research
[4] Department of Medicine, Weill Cornell Medicine, New York, NY, USA.
[5] Novartis Pharma AG, Basel, Switzerland
[6] Biostatistics Research Branch, National Institute of Allergy and Infectious Disease
[7] Department of Biostatistics and Informatics Colorado School of Public Health University of Colorado Anschutz Medical Campus

# Introduction

Real world data (RWD), such as administrative claim records, electronic health records, and large registries, provide unprecedented quantities of data on millions of patients and thousands of variables in real world settings. As such, RWD constitute an extraordinary opportunity to generate practice-based evidence to improve healthcare and health outcomes, so-called real-world evidence (RWE). Recognizing the importance of RWE for regulatory purposes, the United States Congress passed the 21st Century Cures Act1 that mandated the development of United States Food and Drug Administration (FDA) guidance on regulatory use of RWE to support regulatory decisions. Despite the many potential advantages, the prospect of incorrect effect estimates has historically cast doubt on the use of RWE for regulatory science. Indeed, the principle that "correlation does not imply causation" is a fundamental concept used across various scientific fields to prevent logical fallacies and erroneous scientific conclusions, which are rightfully central to most criticisms of using real-world data (RWD) for regulatory science.

However, scientists frequently gain knowledge about cause and effect based on statistical associations. For instance, a statistical association may be interpreted as a causal relationship when it is known that there is no unmeasured confounding, and the direction (e.g., time-ordering) of the causal relationship is already known. One can make such strong assumptions given external knowledge, for example, that data comes from a perfectly executed randomized study with no loss-to-follow-up and perfect adherence. Broadly speaking, causal interpretation must be supported by external knowledge of the data generating process, such as study design or mechanistic knowledge about the phenomena under investigation. This external knowledge is often encoded in a *causal model,* and the set of models and data analysis tools concerned with the appropriateness of such causal interpretations is known as *causal inference* (please see our companion paper on *the causal roadmap* for a more detailed discussion on causal models and causal inference).

Positing causal models with RWD involves making non-testable assumptions, such as assuming the absence of unmeasured confounding variables, time-ordering between the variables, no adjustment for colliders, monotonicity for instrumental variables, etc. Absence of unmeasured confounding is an important assumption that must primarily be addressed at the causal model stage by making every effort to posit a causal model that corresponds to the state-of-the-art in the substantive field, and by making every effort to measure all confounders dictated by the model. For instance, RWD analyses seeking to establish the effectiveness of COVID-19 vaccines for the prevention of Post-Acute Sequelae of COVID (PASC) require understanding and measuring all the patient characteristics that lead patients to get vaccinated in the real world, as well as whether they are likely to affect the risk of developing PASC. However, despite best efforts, there may be situations where the causal model is incorrect, or where some confounders are unmeasurable with current technology or available data. For instance, in an analysis based on Electronic Health Records, certain important socioeconomic factors that may confound the vaccination-PASC relation may be unmeasured. In such cases, the statistical parameter targeted by the analysis may not have a causal interpretation. The use of RWD for regulatory science requires maximum efforts to ensure dependable causal inferences, even when the assumptions of the causal model are incorrect. In the context of plausible violations to the assumptions of the causal model, or the inability to measure some of the confounders dictated by the model, sensitivity analyses are a valuable tool that can be used to make more dependable causal inferences from RWD.

While we often cannot validate an untestable assumption, we can often test how sensitive our scientific conclusions are to violations of our assumptions. To this end, we use a *sensitivity parameter* which encodes the severity of violations to the assumptions of the model, with the goal of determining if the maximum sensible value of the sensitivity parameter (which should be pre-specified, as discussed below) is large enough to invalidate the scientific conclusions derived from adjusted statistical estimates. This simple but powerful idea has a long-standing history in epidemiological sciences, and is currently part of the International Council for Harmonization (ICH) E9 Guidance on Statistical Principles for Clinical Trials.[3,4] One of the most well-known examples is its application in 1959 by Cornfield et al.,[5] who demonstrated that if an unmeasured confounder can explain the observed association between smoking and lung cancer, it would need to cause a nine-fold increase in the probability of smoking. Multiple attempts were made to find such a strong confounder, but all such conjectured confounders (e.g., genetic, hormonal) had an effect on smoking that was much lower than the nine-fold increase necessary to invalidate causal conclusions. As a result, Cornfield et al. concluded that smoking causes lung cancer. This analysis played a pivotal role in establishing a public consensus about the causal relationship between smoking and lung cancer.[6] Others arrived at qualitatively similar conclusions using alternative sensitivity analyses.[7]

The smoking and lung cancer example is a "success" story in the sense that it exemplifies a case where sensitivity analyses prove that an observed association is causal. Perhaps more importantly, sensitivity analyses can be used in the opposite direction to unveil cases where unmeasured confounding could easily explain away an observed association. An example is the effect of hormone replacement therapy (HRT) on cardiovascular disease (CVD), where multiple observational studies showed that HRT reduced the risk of CVD,[8,9] but subsequent randomized trials demonstrated that in fact HRT increases the risk of CVD.[10] If the original observational studies had conducted a sensitivity analysis, they would have found that an unmeasured confounder with a weak association with the exposure (odds ratio 1.13) would have been sufficient to explain away the observed protective association,[11] although it is worth noting that some controversy remains about the effect of HRT.[12]

Before we proceed, it is important to clarify that we refer to sensitivity analyses as methodologies that aid in testing the extent to which varying violations of *causal modeling assumptions* would lead to different conclusions. This kind of sensitivity analysis must be distinguished from analyses that seek to test the extent to which *statistical modeling assumptions* would lead to different conclusions. Statistical and causal sensitivity analyses are fundamentally different in that the former seeks to assess the validity of testable assumptions, whereas the latter seeks to assess the validity of untestable assumptions. For instance, goodness of fit of a logistic regression model may be tested by assessing predictive accuracy after adding additional terms or comparing to other regression models. In contrast, it is impossible to learn from data whether we have measured all the relevant confounders, or whether some of the variables that we are adjusting for are not confounders but are colliders and therefore induce bias. Causal modeling assumptions must be therefore supported based on background substantive information, and when doubted, must be tested with an appropriate sensitivity analysis.

The objective of sensitivity analysis may be simple, but the methods used to express violations of model assumptions, to define sensitivity parameters, and to test their magnitude can be complex. In this article, we provide a brief review of various methods for sensitivity analysis and demonstrate their usefulness in using RWD to establish causality to support regulatory submissions. We begin with a case study that presents an observational analysis of the effectiveness of Nifurtimox (NFX), a medication for the treatment of Chagas disease. We then proceed with a review of the most common methods for sensitivity analysis and conclude with recommendations for their use in supporting regulatory submissions.

# Case study: the effectiveness of Nifurtimox in the treatment of the Chagas disease

## Background on the Chagas disease

American *trypanosomiasis*, also called Chagas disease, is caused by the parasite *Trypanosoma cruzi.*, which is transmitted by an insect vector. The disease affects around 8 to 10 million people in the endemic zones of Latin America, from the South of the United States (US) to the North of Argentina. Although the disease was traditionally restricted to Latin America, a growing number of cases have been reported in the US. Today, the disease is classified as one of the leading neglected tropical diseases in the US,[13] with up to 350,000 persons infected. *T. cruzi* is transmitted by the bite of several species of hematophagous bugs. The parasites are excreted in the feces of the bugs and penetrate human hosts through the mucosa or through scratches in the skin. After localized multiplication, the parasite is then dispersed to target organs (principally the intestinal or cardiac nerve plexus) through invasion of the bloodstream. The acute phase following infection lasts 4-6 weeks and is generally asymptomatic but may lead to fever, malaise, myalgia, and headaches. In more than one third of chronically infected individuals, clinical disease reappears after a period of latency lasting between 10 and 30 years. The chronic stage of the disease manifests as irreversible lesions mainly affecting the cardiac and digestive systems. The chronic form is also associated with a risk of sudden death. Diagnosis is made following detection of trypanosomes in the blood in the acute phase or through serological testing which detects antibodies made to fight the trypanosome infection.[14]

Nifurtimox is one of the drugs currently used in endemic areas of Latin America to treat the Chagas disease. Despite the public importance of the disease, Nifurtimox is currently not approved by the FDA for adults, partly because few research studies exist about its efficacy. Nifurtimox was first approved in the US for the treatment of Chagas disease in pediatric patients on the basis of the results of a randomized study that established the effect of the drug to induce negative seroreversion or seroreduction >= 20% one year after treatment.[15] The long incubation periods of the disease (up to 30 years) means that the cost of a randomized study to assess the effectiveness of Nifurtimox in the full long-term span of the disease is prohibitive.

## Data source

Few studies, randomized or otherwise, exist that follow groups of patients over such long periods of time and provide a proper long-term account of the clinical efficacy of treatment with Nifurtimox. One such study, conducted by Fabbro et al.,[16] followed a group of 404 patients recruited between 1976 and 1999. Data from this study had the following problems which made them not immediately usable for assessing the effectiveness of Nifurtimox:

1. Treatment was assigned mostly based on availability, patients' willingness to be treated, often considering the baseline health status of the patient. Consequently, a naive analysis of the data that does not adjust for these confounders will result in biased inference.

2. Since some patients were lost to follow-up during the study, outcome data are subject to informative missingness. If the reasons why patients were lost to follow-up during the study are related to the outcome of interest (e.g., patients lost to follow-up were because of their health status), ignoring that information will also result in biased inference.

The outcome of interest in this study is negative seroconversion 30 years after treatment (henceforth referred to as seroreversion), meaning that no evidence of presence of the parasite remains in serological blood tests. Due to the long study period, there is substantial loss-to-follow-up. The table below presents the distribution of the outcome across treatment groups in the study.

|  | NFX | Control |
| --- | --- | --- |
| Seroreversion | 16 | 1 |
| No seroreversion | 3 | 27 |
| Lost to follow-up | 36 | 321 |
| Total | 55 | 349 |

The potential for bias is clear from this table. With over 90% of observations lost to follow-up in the control group, it may initially seem impossible to use this data to assess the effectiveness of NFX on 30-year seroreversion without bias.

To overcome the initial barrier of large loss-to-follow-up rates, it is possible to consider external information, such as the small rate of seroreversion when patients are untreated (henceforth referred to as *spontaneous seroreversion*). For instance, two studies in children report a rate of about 5%,[17,18] while a meta-analysis of studies in adults reports a rate as low as 2%.[19] The significance of these low rates becomes clear when compared to the most conservative imputation strategy for the missing NFX patients. If all 36 lost-to-follow-up NFX patients did not serorevert, the resulting NFX seroreversion rate would be 16 out of 55, or 29%. This is considerably higher than the externally supported rate of 5% for spontaneous seroreversion. However, even if the rate of spontaneous seroreversion is as high as 10%, 15%, or 20%, the data still support the hypothesis that NFX induces seroreversion in Chagas patients.

In what follows we use this dataset as an illustrative example for how to conduct a sensitivity analysis, keeping in mind that regulatory decision-making also relies on multiple additional issues such as whether the data are "fit-for-purpose",[20] which we do not address here. The analysis is based on using the rate of spontaneous seroreversion as a *sensitivity parameter,* where the conclusions of effectiveness of NFX are assessed in light of various plausible values of this sensitivity parameter.

## Non-parametric methodology for sensitivity analysis using rates of seroreversion as a sensitivity parameter

The above ideas may be formalized in a rigorous statistical procedure for sensitivity analysis as follows. First, consider a target estimand of interest defined as the *average treatment effect on the treated*, $\psi^c = E[Y(1) - Y(0)| A = 1]$, where $A = 1$ denotes treatment with NFX and $A = 0$ denotes control, $Y(1)$ denotes the potential 30-year seroreversion status of a patient if, possibly contrary to fact, they were treated with NFX, $Y(0)$ denotes the potential 30-year seroreversion status of a patient if, possibly contrary to fact, they were untreated, and $E[Y(1) - Y(0) | A = 1]$ denotes taking the expectation (mean) of the difference between potential outcomes in the population of treated patients. The parameter $\psi^c$ is the *target causal estimand*, interpreted as the difference in outcome rates among treated patients in hypothetical worlds where NFX was given to all vs no Chagas patients. If we knew this number, we would know whether NFX induces higher rates of seroreversion in the patients who are treated with it. Without further assumptions, this quantity is not estimable since we cannot possibly observe a patient's outcome under treatment *and* under no treatment.

In addition to the data on treatment, seroreversion, and loss-to-follow-up, Fabbro et al. collected multiple important baseline variables on the patients in the cohort, including age, sex, initial serology titers, as well as the presence of Chagas-related abnormalities in the electrocardiogram. We use the letter $W$ to denote a vector containing these variables and use $C = 1$ to indicate that a patient had complete follow-up up and the study endpoing was obseved, and $C = 0$ to denote that a patient was lost to follow-up and the endpoint was unobserved. Furthermore, we perform the conservative imputation mentioned above, such that patients treated with NFX who are lost-to-follow-up are assumed to not have seroconverted (death and other potential long-term side effects are not a concern for NFX[21]). This allows us to conservatively approximate $E[Y(1)| A = 1]$ as $E[Y| A = 1]$—the observed outcome rate among the treated. For approximating $E[Y(0)| A = 1]$, if the variables $W$ contain all common causes of treatment, loss-to-follow-up, and outcome, then it can be proved mathematically that

$$E[Y(0) \mid A = 1] = E[E(Y \mid C = 1, A = 0, W) \mid A = 1],$$

where the right-hand side of the above expression can be estimated by running a regression of the outcome on baseline variables among observed controls and using that regression to predict the outcomes that would have been observed for treated patients had they not been treated. This is accomplished by averaging the predicted outcomes over the empirical distribution of W among the treated. Estimators with better performance are also available, we refer the reader to our companion article published in this edition of the journal for a discussion on optimal estimation. This yields a target statistical estimand equal to

$$\psi = E[Y| A = 1] - E[E(Y \mid C = 1, A = 0, W) \mid A = 1],$$

which, contrary to the causal estimand $\psi^c$ is a quantity that can be estimated from data. The fundamental problem is that the assumptions required for establishing the equality $\psi^c = \psi$, namely that $W$ contains all common causes of treatment, loss-to-follow-up, and outcome, is unlikely to hold in this study. We must therefore study the so-called *causal gap,* defined as the difference between the causal target and the statistical target, i.e., $\psi - \psi^c$. In the supplementary materials we show that this causal gap may be bounded as $\psi - \psi^c \leq E[Y(0) \mid A = 1]$. The right-hand side of this inequality is precisely the probability of spontaneous seroreversion that we would have observed for treated patients had they not been treated.

Consider now the null hypothesis of no treatment effect of Nifurtimox, i.e., $\psi^c \leq 0$. According to the above discussion, this hypothesis is true if the hypothesis $\psi \leq E[Y(0) \mid A = 1]$ is true. While the hypothesis $\psi^c \leq 0$ cannot generally be tested, the hypothesis $\psi \leq E[Y(0) \mid A = 1]$ can be tested for varying user-given conjectured levels of the probability of spontaneous seroreversion. If this hypothesis is rejected even for the largest feasible values of the probability of spontaneous seroreversion, then we can be confident that the causal hypothesis of no treatment of NFX may also be rejected.

The probability of spontaneous seroreversion is a *sensitivity parameter*, meaning it's a parameter that is useful for sensitivity analyses. We don't know its true value, but we can make conjectures about plausible values based on our knowledge of the subject matter. It is important that pre-alignment and pre-specification of the range of plausible values occurs prior to the conduct of the analyses.

## Results of sensitivity analysis for the effect of Nifurtimox on the Chagas disease

We analyzed the data of Fabbro et al. using the above sensitivity analysis. The statistical significance of the hypothesis test is given in the figure below as a function of conjectured values for the probability of spontaneous seroreversion. This figure allows us to conclude that, if we believe that the probability of spontaneous seroreversion among the treated is smaller than 0.19, we can reject the hypothesis of no treatment effect of Nifurtimox with a with a two-sided type I error rate of at most 0.05. All the epidemiologic studies as well as biological knowledge about the Chagas disease suggest that the rate of spontaneous seroreversion is smaller than 5%.

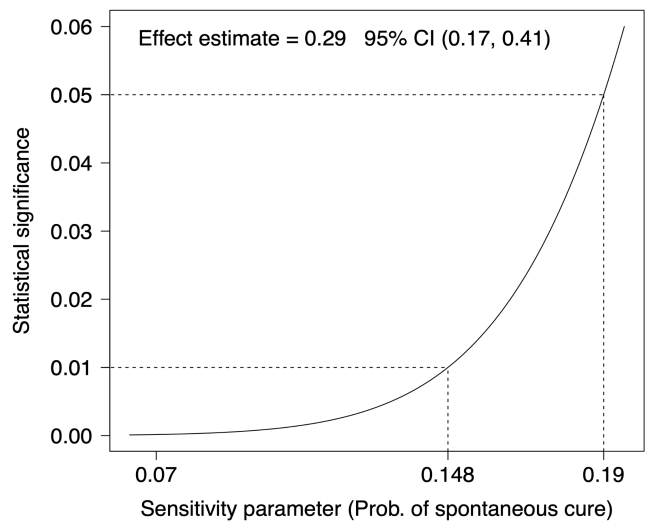

Additional technical details about this sensitivity analysis as well as the methods used to estimate the causal parameter $\psi$ are available in the supplementary materials.

# Current landscape and existing methods for sensitivity analysis

In this section we review some of the most common methods for sensitivity analysis and provide comments on their strengths and weaknesses. This review is not exhaustive, and the reader is referred to Liu et al.[22] and Richardson et al.[23] for more extensive reviews.

## Semiparametric sensitivity analysis

The assumption of no unmeasured confounders may be stated mathematically in multiple ways. One of them is the assumption of independence between the potential outcomes $Y(a)$ and the exposure $A$ of interest (often conditional on observed confounders $W$). The main idea behind semiparametric sensitivity analyses is to posit a model relating the potential outcomes to the exposure of interest.[24–26] For instance, one may posit that the probability of exposure $A = 1$ conditional on potential outcome $Y(a)$ (and possibly covariates $W$) follows a main-terms logistic regression model. The causal effect of $A$ on $Y$ is then identifiable except for the coefficient in front of $Y(a)$ in the above logistic regression. This coefficient, interpreted as the log-odds ratio between $Y(a)$ and $A$, can be used as a sensitivity parameter that quantifies the magnitude of unmeasured confounding. Analysis may therefore proceed by estimating the causal effect for multiple conjectured values of the sensitivity parameter and judging the plausibility of each such value based on subject-matter expert knowledge.

A disadvantage of this approach is that the sensitivity analysis itself requires positing untestable assumptions about a model relating the exposure and the potential outcomes. It is unclear whether misspecification of this model carries serious implications in terms of bias, but it would generally be preferable to rely on sensitivity analyses that do not make extra assumptions. Relatedly, the sensitivity parameter must be informed by subject matter expert knowledge, but it is defined in a scale that is unintelligible and refers to a convenient mathematical construction (e.g., an odds ratio in a logistic regression between $Y(a)$ and $A$) rather than a fundamental property of nature. This makes it hard for subject-matter experts to judge on the plausibility of specific values of the sensitivity parameter.

As an example of this approach, Franks et al.[26] conduct a sensitivity analysis on the effect of antihypertensives on diastolic blood pressure (DBP) using the National Health and Nutrition Examination Survey data. They conclude that, if one is willing to assume that the adjusted odds of receiving antihypertensives in a logistic regression model increases by 1.01 for every additional mmHg in hypothetical counterfactual DBP outcomes under treatment or control, then an otherwise protective but non-significant effect becomes significant. This example illustrates the difficulty in assessing the plausibility of the sensitivity parameter values. Is a logistic regression adjusted odds ratio between counterfactual DBP outcomes and antihypertensives of 1.01 plausible or implausible? The answer to that question depends non-trivially on the variables included in the model as well as on the correctness of the model, which is potentially as difficult to assess as the original "no unmeasured confounder" assumption.

## Nonparametric sensitivity analysis

In contrast to semiparametric sensitivity analyses, nonparametric analyses make no assumptions on the functional form of the relations between variables. This type of sensitivity analysis focuses directly on studying the causal gap with a goal of establishing bounds on it that may be used as sensitivity parameters.

The analysis of the effectiveness of Nifurtimox in the treatment of the Chagas disease presented above is an example of a nonparametric sensitivity analysis. A more general version of this idea has been developed,[19,27] where the goal is to construct bounds on the causal gap using sensitivity parameters that have immediate substantive interpretations, so that the plausibility of their values can be easily judged using a-priori subject-matter knowledge (e.g., the probability of spontaneous seroreversion).

A second example of a nonparametric sensitivity analysis uses E-values[28,29] to posit the existence of an unmeasured confounder $U$, and creates bounds on the causal gap in terms of conjectured magnitudes of the $U \to A$ and $U \to Y$ relations on a risk-ratio scale. These risk ratios are then used as sensitivity parameters. This approach generalizes the sensitivity analysis of Cornfield et al[29–31] in the sense that it seeks to find the minimum effect of an unmeasured confounder such that the observed effect would be completely explained away. As an example of the use of E-values, Bosch et al.[30] recently studied the effectiveness of fludrocortisone and hydrocortisone on death or discharge to hospice in the treatment of patients with septic shock. Their analyses adjusting for measured confounders found a significant absolute risk difference of -3.7% (95% CI -4.2% to -3.1%) comparing hydrocortisone-fludrocortisone vs hydrocortisone alone. Their sensitivity analysis using E-values concluded that an unmeasured confounder that increases the likelihood of treatment and outcome by 37% would be sufficient to explain away the significant effect found in the analyses.

Importantly, E-values cannot accommodate complex high-dimensional confounders. Furthermore, some E-value analyses make strong assumptions, such as assuming that the risk ratio between the unmeasured confounder and the exposure is equal to the risk ratio between the unmeasured confounder and the outcome, as well as the assumption that the prevalence of the uncontrolled confounder among the exposed is 100%.[31] Multiple other methods exist that rely on similar ideas but make parametric assumptions on the $U \rightarrow A$ and $U \rightarrow Y$ relations to incorporate complex confounders,[32–34] although methods relaxing these assumptions also exist.[35–37]

## Identification bounds

Identification bounds are not formally a method for sensitivity analysis in the sense that they do not rely on assessing plausible values for a sensitivity parameter. However, they serve the same purpose of providing information about causal relationships in the presence of unmeasured confounders. The main idea behind identification bounds is to estimate an interval (different from a confidence interval) that bounds the causal effect of a treatment, where this interval is guaranteed to contain the causal effect under no assumptions on the extent of unmeasured confounding.

For example, Bhattacharya et al.[38] used identification bounds to study the effect of right heart catheterization (RHC) on 30-day mortality amongst ICU patients. There is considerable debate in the clinical literature regarding the use of RHC as a diagnostic tool, and its use has been recommended only when there is uncertainty about the best treatment.[39] Therefore, unmeasured confounding is a likely threat to the conclusions of observational analyses of effects of RHC. Using two different types of analyses that allow for any kind of unmeasured confounding, Bhattacharya et al. found that RHC had either a null or a protective effect on 30-day mortality, whereas prior studies that assumed no unmeasured confounders had found RHC to increase 30-day mortality.[40] Although the analyses of Bhattacharya et al. rely on an instrumental variable assumption, multiple identification bounds in the literature do not require this or any other assumption.[41]

Identification bounds are most commonly used in the econometrics literature, but they have also been used to assess the comparative effectiveness of treatments in RWD, as illustrated by the above example. Because it relies on few assumptions and has an ambitious goal, this methodology sometimes results in wide bounds that may be uninformative. Manski[41] and Molinari[42] provide a comprehensive review of existing methods for identification bounds.

## Negative controls

Additional methods such as negative control treatments and outcomes may be used to rule out the possibility that observed adjusted associations are due to unobserved confounding.[43] For instance, Dickerman et al.[44] recently used RWD to assess the comparative effectiveness of COVID-19 vaccines in a real world population of US veterans. It is thought that COVID-19 vaccines cannot possibly influence infection status in the 10-day period following the vaccination. Thus, infection status at 10-day post vaccination may be used as a negative control outcome. Specifically, if a procedure purported to estimate causal effects yields a non-null effect on this outcome, then that procedure must be ruled out as giving biased causal estimates. A review by Shi et al.[45] provides further examples of successful use of negative controls in applied research. This kind of ad-hoc negative control does not guarantee that an association may be interpreted causally but can be used to rule out non-causal associations, although recent efforts have been made in the statistics literature to formalize the use of negative controls for identification of causal parameters in the presence of unmeasured confounding.[46,47]

Table 1 below summarizes the assumptions, advantages, and disadvantages of the above types of sensitivity analysis.

| Type of sensitivity analysis | Disadvantages | Advantages | Example |
|---|---|---|---|
| Semiparametric | • Requires arbitrary models for unmeasured variables.<br>• Requires positing plausible values for the unintelligible coefficients of the above model. | • Mathematical convenience. | • Franks et al.[43]—antihypertensives and diastolic blood pressure. |
| Nonparametric | • Typically, none; although some methods such as instances of the E-value might use implausible assumptions.[31] | • Requires positing plausible values for intelligible scientific quantities (e.g., spontaneous probability of Chagas seroreversion). | • Nifurtimox on Chagas disease, discussed in this manuscript. |
| Negative controls | • Does not conclusively guarantee that associations are causal. | • Requires positing an outcome with a null treatment effect, which is often feasible. | • Dickerman et al.[48]—COVID vaccines |
| Identification bounds | • Bounds are often too wide to be informative. | • Operates with few or no assumptions. | • Bhattacharya et al.[38]—right heart catheterization and 30-day mortality |

Table 1 Types of sensitivity analyses described and their advantages and disadvantages.

## Sensitivity analysis considerations when using RWD for regulatory science.

### Prespecification

Prespecification of a study refers to publication in complete detail of the study design and analysis plan before all data are collected and analyses are conducted.[48] As with all aspects of a data analysis, sensitivity analyses must be fully prespecified to appropriately control type I error and avoid biases due to researcher degrees of freedom.[49] FDA guidance does allow for choices to be made using blinded data with pre-specification of the plan after such examination.[50] Prespecification of the analysis must include the range of plausible values for the sensitivity parameter, which may be based on prior literature or consensus in the substantive field. For instance, a prespecified analysis plan for the case study of the effect of Nifurtimox on the Chagas disease may have conservatively prespecified 10% as the maximum possible rate of spontaneous seroreversion among the treated, based on prior literature that suggests that this rate is of at most 5%.[16,17-18]

The need for prespecification means that it is important that the sensitivity parameters used have an interpretation that corresponds to interpretable phenomena rather than convenient mathematical formalizations, as in our illustration on the effect of NFX on the Chagas disease. This ensures that prespecified plausible values may be obtained through consultation with experts or the literature. The need for prespecification makes it harder to use sensitivity parameters interpreted as the coefficient relating the exposure $A$ and the potential outcome $Y(a)$ in a logistic or linear regression model. Likewise, analyses that rely on a sensitivity parameter interpreted in terms of the strength of the associations $U \rightarrow A$ and $U \rightarrow Y$ will usually require that the unmeasured confounder $U$ is specified and described in terms of real-world phenomena, even if it is not possible to measure it. Arbitrary unspecified confounders will make it difficult for subject matter experts to obtain prior information that can inform plausible values for the associations $U \rightarrow A$ and $U \rightarrow Y$.

## Sensitivity analyses with assumptions

Some methods for sensitivity analysis use statistical models to obtain mathematical expressions of violations of the assumptions of the model. For example, a strand of the literature makes the assumption that the probability of treatment $A$ within strata of the potential outcome $Y(a)$ (and possibly measured confounders) follows a logistic regression model.[23–25] Other methods directly assume statistical models that capture the dependence between the outcome $Y$ and a hypothetical unmeasured confounder $U$, for example assuming that they are linearly related.[33] Like all statistical models, these models are subject to misspecification. For instance, it could be the case that the relation between the confounder $U$ and the outcome $Y$ is quadratic, so that a linear approximation will fail to account for unmeasured confounding. Unlike statistical models applied to real data, models for unobserved variables such as $Y(a)$ and $U$ are not testable. Therefore, while using sensitivity analyses based on models is certainly better than not performing a sensitivity analysis at all, it is preferable to use sensitivity analyses that make no assumptions about the mathematical nature of the unmeasured confounding.

## Summary and conclusions

Sensitivity analysis is an important tool that can help researchers test whether causal conclusions obtained from analyses of observational data are robust to violations of assumptions of the causal model. The routine use of sensitivity analyses with RWD increases the trustworthiness of effectiveness conclusions for regulatory science. Sensitivity analyses are most likely to be useful and informative when other aspects of the study (described in our companion paper on the *causal roadmap*) are also carefully designed. That is, sensitivity analyses on their own are not a panacea and cannot save a poorly designed and conducted analysis of RWD. As with other study aspects, pre-specification of sensitivity analyses for RWD in regulatory settings is crucial to avoid wrong conclusions due to researcher degrees of freedom.[51] Most sensitivity analyses require auxiliary scientific information (e.g., the probability of spontaneous seroreversion in the Chagas disease example discussed) to produce meaningful conclusions, although some methods such as those based on identification bounds can sometimes produce meaningful conclusions without such knowledge.

In this paper we focused on an illustration of sensitivity analyses for the assumption of unmeasured confounding, but causal models often entail other important assumptions which may also be subject to sensitivity analysis.

There is a vast literature on sensitivity analysis for causal inference with many fields contributing distinct approaches and tools . For instance, the computer science and machine learning community has developed software tools such as PyWhy[52] that help scientists capture causal assumptions and apply sensitivity analyses and other refutations. Furthermore, there are numerous developed and emerging methods that rely on different assumptions appropriate to a variety of scenarios, such as the identification of secondary small-scale or continuous experiments to infer or validate causal assumptions (i.e., adaptive, active sampling, or reinforcement learning);[53,54] and explorations of large language models as a source of domain knowledge for semi-automated critiquing and refinement of researchers' causal assumptions.[55] Such tools and emerging methods and their requirements should be considered and assessed carefully before use in regulatory science. In all cases, it is important to specify sensitivity analysis that have at least two important properties: (i) their conclusions do not rely on further untestable assumptions, and (ii) the sensitivity parameter has a clear scientific interpretation so that pre-specification of a plausible range of values is possible from available subject-matter knowledge.

## Acknowledgements

We would like to thank the sponsors of the FIORD workshop, including the Forum for Collaborative Research and the Center for Targeted Machine Learning and Causal Inference (both at the School of Public Health at the University of California, Berkeley), and the Joint Initiative for Causal Inference. We would also like to thank MetronomX for providing access to the dataset used in the paper, as well as Lauren Dang and Alexander D'Amour for useful discussions.

# Supplement

Data structure and notation

The objective of this data analysis will be to estimate the causal effect of the treatment on the probability of developing a medical outcome 30 years after enrollment. The eligible medical records of the patients were reviewed, and the following information was extracted:

- Baseline variables $W$:
    - Patient age at baseline
    - Sex
    - Date of first visit
    - Xenodiagnosis at baseline (if conducted)
    - Initial serology
    - Physical examination and ECG tests within first 6 monthd from first visit
- Treatment $A$, with levels nifurtimox, benznidazole or control
- Indicator $C$ of lost to follow-up. The criteria for lost to follow up are:
    - Patients who expressed not to be followed up
    - Patients who could not be reached to be followed up
    - Patients who moved outside Santa Fe
    - Patients who traveled to endemic areas
    - Patients who became unable to attend medical examinations at the center

    A patient's outcome will be considered missing ($C = 0$) if he/she was lost to follow-up during the study at any time point before $t$ years have passed, even if the medical outcome of interest has already occurred.
- Primary outcome $Y$, an indicator of seroconversion having occurred before time $t$. This outcome is subject to missingness because some patients were lost to follow-up during the study.
- Secondary outcome $Z$, an indicator of not having developed Chagas related ECG abnormalities before time $t$. This outcome is subject to missingness because some patients were lost to follow-up during the study.

The data of interest can be coded in terms of the following data structure:
$$O = (W, A, C, Y, Z). \qquad (1)$$

To define the causal parameter of interest we use state-of-the-art methods for causal inference. Consider the following nonparametric structural equation model (NPSEM Pearl, 2000):
$$\begin{aligned} W &= f_W(U_W) \\ A &= f_A(W, U_A) \\ C &= f_C(W, A, U_C) \\ Y &= C f_Y(W, A, U_Y) \\ Z &= C f_Z(W, A, Y, U_Z) \end{aligned} \qquad (2)$$

which allows the definition of the counterfactual outcomes $Z_a, Y_a : a \in \{0, 1\}$ corresponding to the outcomes in an intervened NPSEM in which the equations $f_A$ and $f_C$ are removed and $A = a,\ C = 1$ are set deterministically. Thus, $Y_a$ represents the outcome of a subject if, possibly contrary to the fact, a subject would have received level $a$ of treatment and censoring was not present.

Analysis for the primary endpoint of seroreversion

As previously mentioned, causal effects are defined in terms of the distribution of the counterfactual outcomes $Y_a$. For example, $E(Y_1|A=1)$ denotes the cure rate among the treated if, contrary to the fact, we had treated and observed the outcome of all patients. The objective of this analysis is to estimate.

$$\psi_0^f = E(Y_1 - Y_0|A=1),$$

the causal effect of treatment among the treated. Under the randomization assumption that

$$(U_A, U_C) \perp\!\!\!\perp U_Y|W, \qquad (3)$$

we have

$$E(Y_a|A=1) = E_W\{E(Y|C=1, A=a, W)|A=1\},$$

where the right-hand side only can be estimated from the observed data alone. However, since $W$ may not contain all the confounding factors that determine missingness and treatment allocation, it is possible that the randomization assumption (3) does not hold. Therefore, we will perform a sensitivity analysis to measure the extent to which the efficacy of Nifurtimox can be established even under violations to this randomization assumption.

Sensitivity analysis

This sensitivity analysis was first proposed by Díaz and van der Laan (2012). First, we will impute all the treated patients that were missing as treatment failures. As we will see, this provides a very conservative scenario for estimation of the causal effect of nifurtimox. The reason to use this conservative approach is that it yields a sensitivity parameter that is easy to interpret. Denote this imputed outcome by $Y^*$.

We will now address the problem of estimating $E(Y_a|A=1)$ for $a \in \{0,1\}$. Firstly, the counterfactual expectation $E(Y_1|A=1)$ can be conservatively approximated by $E(Y^*|A=1)$. That is

$$E(Y^*|A=1) = E(CY_1|A=1) \leq E(Y_1|A=1). \qquad (4)$$

As we will see, the use of this approximation will yield a sensitivity parameter with a more straightforward interpretation. Secondly, as previously mentioned, under violations to the randomization assumption (3) $E(Y_0|A=1)$ is not equal to $\varphi_0 = E_W\{E(Y|C=1, A=0, W)|A=1\}$. However, we use $\varphi_0$ as an approximation to $E(Y_0|A=1)$, and perform a sensitivity analysis on the amount of bias of this approximation. The value that best approximates $\psi_0^f$ is thus given by the estimand

$$\psi_0 = E(Y^*|A=1) - E_W\{E(Y|C=1, A=0, W)|A=1\}. \qquad (5)$$

Note that, using (4), the bias of this approximation of the desired causal effect $\psi_0^f$ is bounded above by

$$\begin{aligned}\psi_0 - \psi_0^f &= E(Y^*|A=1) - E_W\{E(Y|C=1, A=0, W)|A=1\} \\ &\quad - E(Y_1|A=1) + E(Y_0|A=1) \\ &\leq E(Y_0|A=1) - E_W\{E(Y|C=1, A=0, W)|A=1\} \\ &\equiv \delta_0.\end{aligned}$$

We use $\delta_0$ as a sensitivity parameter to find the values of the bias $\delta_0$ for which the hypothesis of no treatment effect $H_0 : \psi_0^f \leq 0$ would be rejected. An important feature of the sensitivity parameter $\delta_0$ is that under assumption (3) $\delta_0 = 0$. Note also that

$$\psi_0 - \delta_0 \leq \psi_0^f,$$

and therefore, for a given value of $\delta_0$, the hypothesis of no efficacy of nifurtimox $H_0^f : \psi_0^f \leq 0$ implies the hypothesis $H_0 : \psi_0 - \delta_0 \leq 0$. Thus, rejecting $H_0$ implies rejecting $H_0^f$. For given $\delta_0$, if $\psi_n$ is an asymptotically linear estimate of $\psi_0$ with standard error $\sigma_n$, we can use the statistic

$$T = \frac{\psi_n - \delta_0}{\sigma_n},$$

to define a valid test for $H_0^f$ as "*Reject $H_0^f$ if $t > z_\alpha + \delta_0/\sigma_n$*", where $t$ is the observed value of $T$. Equivalently, given the value $t$, we reject $H_0^f$ when the true $\delta_0$ satisfies

$$\delta_0 = E(Y_0|A=1) - E_W\{E(Y|C=1, A=0, W)|A=1\} \leq \sigma_n(t - z_\alpha).$$

We can now plot the upper bound on $\delta_0$ that would lead to a rejection of $H_0^f$ as a function of the desired probability of type I error $\alpha$, and decide on the plausibility of each value of $\delta_0$ based on subject-matter expert knowledge.

A targeted maximum likelihood estimator (van der Laan and Rubin, 2006; Rose and van der Laan, 2011) of $\psi_0$ and its variance estimation are discussed below.

Targeted minimum loss-based estimation (TMLE) of the observed data parameters

The likelihood of the random variable $O$ in (1) can be factorized according to the previous time ordering as:

$$P(O) = Q_W(W)g(A|W)\phi(C|A,W)\{I(C=1)\bar{Q}_Y(Y|C,A,W)\bar{Q}_Z(Z|Y,C,A,W)\}, \quad (9)$$

where $Q_W$ is the marginal distribution of $W$, $g$ is the distribution of $A$ given $W$ (under randomization $A$ would have been independent of $W$), $\phi$ denotes the missingness mechanism given $A$ and $W$ (under complete MAR $C$ would have been independent of $(A, W)$), $Q_Y$ is the seroconversion outcome distribution conditional on $(C, A, W)$, and $Q_Z$ is the Chagas related ECG abnormalities outcome distribution conditional on $(Y, C, A, W)$. We will denote the expectation of $Y$ conditional on $(C, A, W)$ by $\bar{Q}_Y(C, A, W)$, the expectation of $Z$ conditional on $(C, A, W)$ by $\bar{Q}_Z(C, A, W)$, and the expectation of $Z$ conditional on $(Y=0, C, A, W)$ by $\bar{Q}_{Z,0}(C, A, W)$. Estimators of all the relevant quantities in the likelihood (9) will be performed through Super Learner (van der Laan et al., 2007), a machine learning algorithm that will be described in **Error! Reference source not found.**. The TML estimator that we present is discussed in more detail by van der Laan (2010) and Hubbard et al. (2011).

TMLE of parameter (5) for the analysis of seroconversion

Recall the definition of estimand of interest:

$$\psi_0 = E(Y^*|A=1) - E_W\{E(Y|C=1, A=0, W)|A=1\}.$$

The expectation $E(Y^*|A=1)$ will be estimated with its empirical counterpart, and a TMLE will be used to estimate $\varphi_0 = E_W\{E(Y|C=1, A=0, W)|A=1\}$.

Implementation of the TMLE will be performed following the procedure described in chapter 8 of Rose and van der Laan (2011) as follows.

1. Find initial estimators $\bar{Q}_{Y,n}$, $g_n$, and $\phi_n$ of $\bar{Q}_Y$, $g$, and $\phi$.
2. Denote $n_1$ the number of treated patients. For each subject $i$ compute

$$\varphi_n = \frac{1}{n_1} \sum_{i=1}^{n} I(A_i = 1)\bar{Q}_{Y,n}(1,0,W_i)$$

$$L_{1i} = \frac{C_i}{\phi_n(1|A_i,W_i)} \frac{I(A_i=0)g_n(1|W_i)}{g_n(0|W_i)}$$

$$L_{2i} = \bar{Q}_n(1,0,W_i) - \varphi_n,$$

where $i = 1, \ldots, 492$.

3. Estimate $\epsilon_1$ and $\epsilon_2$ in the univariate logistic regression models

$$\text{logit } \bar{Q}_Y(C,A,W) = \text{logit } Q_{Y,n}(A,C,W) + \epsilon_1 L_1$$

$$\text{logit } g(A|W) = \text{logit } g_n(A|W) + \epsilon_2 L_2$$

4. Compute

$$\text{logit } \bar{Q}^*_{Y,n}(C_i,A_i,W_i) = \text{logit } Q_{Y,n}(A_i,C_i,W_i) + \epsilon_{1,n} L_1$$

$$\text{logit } g^*_n(A_i|W_i) = \text{logit } g_n(A_i|W_i) + \epsilon_{2,n} L_{2i}$$

5. Update $\bar{Q}_{Y,n} = \bar{Q}^*_{Y,n}$ and $g_n = g^*_n$

6. Repeat 2-5 until convergence of $\varphi_n$.

The TMLE of $\varphi_0$ is defined as the last value of this iterating process. In a slight abuse of notation, we will denote this converging value by $\varphi_n$. Assuming that the outcome, treatment, and censoring mechanisms are estimated consistently, under additional conditions explained in Appendix 18 of Rose and van der Laan (2011), $\varphi_n$ is an asymptotically linear estimator with known influence curve $D_\psi$:

$$\varphi_n - \varphi_0 = \frac{1}{n} \sum_{i=1}^{n} D_\varphi(O_i) + o_P(1/\sqrt{n}),$$

where

$$D_\varphi(O) = \frac{C}{\phi(1|A,W)} \frac{I(A=0)g(1|W)}{P(A=1)g(0|W)}\{Y^* - \bar{Q}(C,A,W)\} + \frac{I(A=1)}{P(A=1)}\{Q(1,0,W) - \varphi_0\}$$

Thus, the corresponding influence curve of the estimator $\psi_n$ of $\psi_0$ equals

$$D_\psi(O) = -\frac{C}{\phi(1|A,W)} \frac{I(A=0)g(1|W)}{P(A=1)g(0|W)}\{Y^* - \bar{Q}(C,A,W)\} + \frac{I(A=1)}{P(A=1)}\{Y^* - Q(1,0,W) - \psi_0\},$$

and the variance of $\psi_n$ can be estimated as

$$\sigma_n = \frac{1}{n^2} \sum_{i=1}^{n} D_{\psi,n}(O_i)^2$$

where $D_{\psi,n}$ is the plug-in estimate of $D_\psi$.

Super learning

As explained in the previous subsection, the TMLE requires initial estimators of $g$ (treatment mechanism), $\phi$ (censoring mechanism), and $\bar{Q}$ (outcome mechanism). We will use an a priori-specified super learner for each one of these. Super learner is an ensemble learner that finds an optimal convex combination of a list of user-supplied estimators, based on cross-validation using the appropriate loss function (i.e, log-likelihood). It is optimal in the sense that it performs asymptotically as well as an oracle selector based on knowledge of the true distribution of the data. The finite sample size as well asymptotic properties of the super learner have been studied by van der Laan et al. (2007); van der Vaart et al. (2006), among others. The algorithms (Fixed before obtaining the data, immutable from then on) will be given by:

- Logistic regression with main terms
- Boosted logistic regression
- L1 regularized logistic regression
- Bayesian logistic regression with non-informative priors
- Generalized additive logistic regression models using smoothing splines with various degrees of freedom
- Boosted generalized additive logistic regression
- Sample mean

We will briefly describe the Super Learner algorithm; a complete description can be found in van der Laan et al. (2007). Consider the usual setting in which we observe $n$ identically distributed copies $O_i$, $i = 1, \ldots, n$ of the random variable $O \sim P_0$. Super learner deals with estimation of parameters $\theta_0$ defined as the minimizer of the expectation of a loss function $L(O, \theta)$, the so-called risk of $\theta$, over some parameter space $\Theta$. This is $\theta_0 = \arg\min_{\theta \in \Theta} E_0 L(O, \theta)$. Binary regression problems such as the case of the parameters $Q_Y, Q_Z, Q_{Z,0}, g$, and $\phi$ defined in the previous sections can be defined in these terms by using the logistic log-likelihood loss function.

An algorithm estimator $\hat{\Theta}$ of $\theta_0$ can be seen as a mapping that takes the empirical distribution $P_n$ and maps it into an estimate. $\hat{\Theta}(P_n)$ is then the estimator based on the entire sample, and its conditional risk is given by

$$R(\hat{\Theta}, P_0) = \int L\{o, \hat{\Theta}(P_n)\} dP_0(o).$$

This conditional risk of an estimator depends on $P_0$, and is therefore an unknown quantity that needs to be estimated. A first option is to use a plug-in estimator in which $P_n$ is used instead of $P_0$. If the space $\Theta$ is very large, this plug-in estimator of the risk will favor estimators $\hat{\Theta}$ that over-fit the data. Instead, super learner provides an algorithm that uses a v-fold cross validated risk estimate to choose the best estimator of $\theta_0$.

Let $s \in \{1, \ldots, S\}$ index a random sample split into a validation sample $V(s) \subset \{1, \ldots, n\}$ and a training sample $T(s) = \{V(s)\}^c$. Here we note that the union of the validation samples equals the total sample: $\cup_{s=1}^{S} V(s) = \{1, \ldots, n\}$, and the validations samples are disjoint: $V(s_1) \cap V(s_2) = \emptyset$ for $s_1 \neq s_2$. Let $P_{T(s)}$ be the empirical distribution of the training sample $s$, and let $P_{V(s)}$ the empirical distribution of validation sample $s$. The cross validated estimator of the risk is given by the following expression, in which the parameter is estimated on a training set and the risk is estimated in the corresponding validation set:

$$\frac{1}{S} \sum_{s=1}^{S} R\{\hat{\Theta}(P_{T(s)}), P_{V(n)}\} = \frac{1}{S} \sum_{s=1}^{S} \int L\{o, \hat{\Theta}(P_{T(s)})\} dP_{V(s)}(o). \tag{10}$$

Assume that we have a list of candidate estimators $\hat{\Theta}_j : j \in J$. The discrete super learner is defined as the estimator in this list for which the cross validated risk in (10) is the smallest. Consider now a library of candidate estimators given by all possible convex linear combinations of the candidates $\hat{\Theta}_j$. It can be shown (van der Laan et al., 2007) that the candidate in this library with the smallest cross validated risk is be given by

$$\hat{\Theta}(P_n)(O) = \sum_{j \in J} \beta_j \hat{\Theta}_j(P_n)(O),$$

where

$$\beta = (\beta_1, \ldots, \beta_J) = \arg\min_{\beta} \frac{1}{S} \sum_{s=1}^{S} \frac{1}{n_s} \sum_{i \in V(s)} L\left\{O_i, \sum_{j \in J} \beta_j \hat{\Theta}_j(P_{T(s)})\right\}, \quad (11)$$

subject to $\sum_{j \in J} \beta_j = 1$ and $\beta_j \geq 0$ for all $j \in J$. Here $n_s$ denotes the size of the validation sample $s$.